\documentclass[11pt]{article}
\usepackage{amssymb}
\usepackage{amsmath}
\usepackage{graphics}
\usepackage{epsfig}
\usepackage{a4wide}
\usepackage{cite}
\usepackage{multirow}
\usepackage{color}
\usepackage{mathrsfs}

\DeclareGraphicsExtensions{.eps}
\begin{document}

\title{ NLO QCD and electroweak corrections to $ZZ+{\rm jet}$ production with $Z$-boson leptonic decays at LHC}
\author{ Wang Yong$^1$, Zhang Ren-You$^1$, Ma Wen-Gan$^1$, Li Xiao-Zhou$^1$, Guo Lei$^2$ \\
{\small $^1$ Department of Modern Physics, University of Science and Technology of China (USTC),}  \\
{\small  Hefei 230026, Anhui, People's Republic of China} \\
{\small $^2$ Department of Physics, Chongqing University, Chongqing 401331, People's Republic of China}  }

\date{}
\maketitle
\vskip 15mm
\begin{abstract}
In this paper we present the full NLO QCD + NLO EW corrections to the $Z$-boson pair production in association with a hard jet at the LHC. The subsequent $Z$-boson leptonic decays are included by adopting both the naive NWA and MadSpin methods for comparison. Since the $ZZ+{\rm jet}$ production is an important background for single Higgs boson production and new physics search at hadron colliders, the theoretical predictions with high accuracy for the hadronic production of $ZZ+{\rm jet}$ are necessary. We present the numerical results of the integrated cross section and various kinematic distributions of final particles, and conclude that it is necessary to take into account the spin correlation and finite width effects from the $Z$-boson leptonic decays. We also find that the NLO EW correction is quantitatively nonnegligible in matching the experimental accuracy at the LHC, particularly is significant in high transverse momentum region.
\end{abstract}


\vfill \eject
\baselineskip=0.32in
\makeatletter      
\@addtoreset{equation}{section}
\makeatother       
\vskip 5mm
\renewcommand{\theequation}{\arabic{section}.\arabic{equation}}
\renewcommand{\thesection}{\Roman{section}.}
\newcommand{\nb}{\nonumber}

\vskip 5mm
\section{Introduction}
\par
The weak gauge boson pair production with subsequent leptonic decays plays an essential role in the research of Higgs particle and the new physics beyond the standard model (SM). It is always accompanied by additional one or more hard QCD radiations at the Large Hadron Collider (LHC), therefore, the study of $VV^{\prime} + {\rm jet(s)}$ $(VV^{\prime}=WW,ZZ,ZW)$ productions at the LHC is significantly improtant and may also help us to gain a deeper understanding of jet physics. By far the $VV^{\prime}+{\rm jet}$ production cross sections are known up to the QCD next-to-leading order (NLO), and the precision calculations for the NLO QCD + NLO electroweak (EW) corrections are urgently needed to match the experimental accuracy at the LHC. These research items are listed in the 2013 Les Houches high precision wish list \cite{wishlist}.

\par
The $Z$-boson pair production is of great phenomenological importance in measuring gauge couplings at the LHC, and is also one of the background processes for single Higgs boson production. Thus a thorough understanding of the $Z$-boson pair production is necessary. So far physicists have made enormous efforts in investigating $Z$-boson pair production at hadron colliders not only experimentally (see Refs.\cite{exper1,exper2,exper3,exper4,exper5} and more references therein), but also theoretically. The NLO QCD predictions for the $Z$-boson pair production with leptonic decays were calculated in Refs.\cite{ZZ-QCD1,ZZ-QCD2,ZZ-QCD3}, and the next-to-next-to-leading order QCD calculations have been given in Refs.\cite{ZZ-NNLO-QCD-1,ZZ-NNLO-QCD-2} including the significant loop-induced gluon-fusion contributions \cite{gluon-fusion1,gluon-fusion2,gluon-fusion3}. The NLO EW corrections to the $Z$-boson pair production were given in Refs.\cite{ZZ-NLO-EW1,ZZ-NLO-EW2} and then were extended to include leptonic decays with spin correlation effect in Ref.\cite{ZZ-NLO-EW3}. More recently, the investigation to the four lepton production including the full off-shell contributions from the intermediate $Z$-bosons and photons was presented in Ref.\cite{ZZ-NLO-EW4}. The $Z$-pair production is also particularly interesting in searching for new physics \cite{ZZ-BSM1,ZZ-BSM2}, since there is no $ZZ\gamma$ or $ZZZ$ coupling \cite{ZZZ-coupling1,ZZZ-coupling2,ZZZ-coupling3,ZZZ-coupling4} in the SM. Probing such anomalous neutral gauge boson couplings at hadron colliders has also been studied in the literature \cite{ZZZ-anomalous1,ZZZ-anomalous2,ZZZ-anomalous3}.

\par
The $Z$-boson pair production at a hadron collider is always associated with one or more additional hard jets. The complete NLO QCD calculation for the $ZZ+{\rm jet}$ production without $Z$-boson decays at the Tevatron and the LHC has been presented in Ref.\cite{zzj-qcd}, while the precision study including the NLO QCD + NLO EW corrections on the $ZZ+{\rm jet}$ production at hadron colliders with subsequent vector boson decays is still desired \cite{wishlist}. In this work we report on the NLO QCD + NLO EW calculations for the $ZZ+{\rm jet}$ production with $Z$-boson leptonic decays in the SM at hadron colliders. The rest of this paper is organized as follows: The calculation strategy is described in Section 2. Numerical results of the integrated cross section and various kinematic distributions are presented in Section 3, and finally we give a short summary in Section 4.

\vskip 5mm
\section{Calculation strategy}
\subsection{General description}
\label{general-description}
\par
The calculation method for the NLO QCD corrections to the $pp \to ZZ+{\rm jet}$ process is the same as in Ref.\cite{zzj-qcd}. In this section we describe mainly on the differences compared to that paper. In both the NLO QCD and NLO EW calculations we apply {\sc FeynArts}-3.7 \cite{feynarts} to generate the Feynman diagrams and {\sc FormCalc}-7.3 \cite{formcalc} to algebraically simplify the corresponding amplitudes. To check the correctness of our NLO QCD calculation, we perform the NLO QCD calculation by using both {\sc MadGraph}5 \cite{madgraph} and {\sc FeynArts}-3.7+{\sc FormCalc}-7.3+{\sc LoopTools}-2.8 \cite{feynarts,formcalc,looptools}, and find that the numerical results obtained from the two packages are coincident with each other within the calculation errors.

\par
At the leading order (LO), the $pp \rightarrow ZZ + {\rm jet} + X$ process involving the following partonic processes:
\begin{eqnarray}
\label{qqbar+qg}
q \bar{q} \rightarrow ZZg,~~~~~~~
qg \rightarrow ZZq,~~~~~~~
\bar{q} g \rightarrow ZZ\bar{q}.
\end{eqnarray}
In initial state parton convolution we adopt the 5-flavor scheme, i.e., $q = u, d, c, s, b$, and neglect their quark masses. We accomplish our calculation in the 't Hooft-Feynman gauge. In the NLO calculations, the ultraviolet (UV) and infrared (IR) singularities are isolated in dimensional regularization scheme, where the dimensions of spinor and space-time manifolds are extended to $D=4 - 2\epsilon$.

\par
\begin{figure}[htbp]
  \begin{center}
     \includegraphics[scale=0.85]{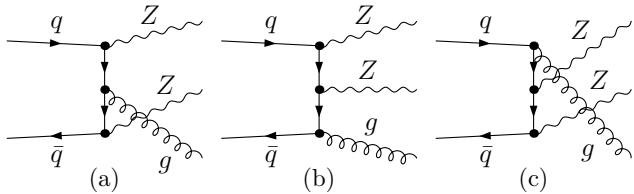}
    \caption{Representative LO Feynman diagrams for the partonic process $q\bar{q} \rightarrow ZZg$.}
   \label{fig1}
  \end{center}
\end{figure}

Some representative LO Feynman diagrams for the subprocess $q\bar{q} \rightarrow ZZg$ are shown in Fig.\ref{fig1}. We can see that the Feynman diagram structure and the helicity amplitude are obviously distinct from those for the $W^+W^-+{\rm jet}$ production \cite{zzj-qcd}. Due to the identical $Z$-boson final state, we encounter considerably increased number of Feynman diagrams in this work compared to the $W^+W^-+{\rm jet}$ production. Meanwhile, the triple gauge boson vertices are involved in the $W^+W^-+{\rm jet}$ production but not in the $ZZ+{\rm jet}$ production. It is noteworthy that there is no need to distinguish the $ZZ+{\rm jet}$ and $ZZ+\gamma$ events in the LO calculation, while we should properly define the $ZZ+{\rm jet}$ event in the NLO EW calculation due to a possible additional photon in final state. This issue will be detailed in Section \ref{EISS}.

\par
\begin{figure}[htbp]
  \begin{center}
     \includegraphics[scale=0.85]{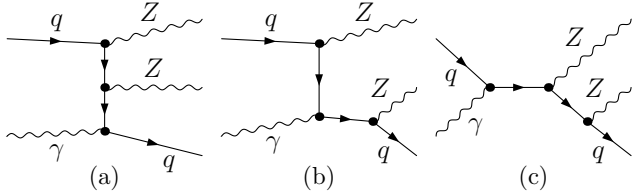}
    \caption{ Representative tree-level Feynman diagrams for the photon-induced subprocess $\gamma q \rightarrow ZZq$.  }
   \label{fig2}
  \end{center}
\end{figure}

The photon-induced subprocesses,
\begin{eqnarray}
\label{gamma-induced}
\gamma q \rightarrow ZZ q,~~~~~~~~\gamma \bar{q} \rightarrow ZZ \bar{q},
\end{eqnarray}
also contribute to the parent process $pp \rightarrow ZZ + {\rm jet} + X$. In Fig.\ref{fig2} we present the representative tree-level Feynman diagrams for the $\gamma q \rightarrow ZZq$ partonic process. Since the NLO QCD corrections to the photon-induced subprocesses are at ${\cal O}(\alpha^3 \alpha_s)$, the same order as the EW corrections to the quark-antiquark annihilation and quark-gluon fusion subprocesses, we should also include these photon-induced contributions in the calculation of the NLO QCD+EW corrections to the $ZZ + {\rm jet}$ production.

\par
Now we clarify the appropriate choice of input fine structure constant $\alpha$ in this work. The renormalized electric charge is given by
\begin{eqnarray}
e_{0} = ( 1 + \delta{Z_e} ) e,
\end{eqnarray}
where $e_0$ is the bare electric charge and $\delta Z_e$ is the corresponding renormalization constant. In the $\alpha(0)$-scheme, the on-shell renormalization condition is employed for the $e-e-\gamma$ vertex in the Thomson limit (photon momentum transfer $k^2 \to 0$). Then we get the electric charge renormalization constant as \cite{alpha0-input}
\begin{eqnarray}
\delta{Z^{\alpha(0)}_e} = -\frac{1}{2} \delta{Z_{AA}} - \frac{\sin\theta_W}{\cos\theta_{W}} \frac{1}{2} \delta{Z_{ZA}},
\end{eqnarray}
where the wave-function renormalization constants $\delta{Z_{AA}}$ and $\delta{Z_{ZA}}$ are given by
\begin{eqnarray}
\delta{Z_{AA}} = -\frac{\partial{\sum_{T}^{AA}(k^2)}}{\partial{k^2}}{\Big|}_{k^2 \rightarrow 0},~~~~~~~~~~
\delta{Z_{ZA}} = 2\frac{\sum_{T}^{AZ}(0)}{M_{Z}^2}.
\end{eqnarray}
$\sum_T^{XY}(k^2)$ is the transverse part of the unrenormalized self-energy of the $X \rightarrow {Y}$ transition at momentum squared $k^2$. As we know, the electric charge renormalization constant $\delta{Z^{\alpha(0)}_e}$ contains mass-singular terms $\log(m_f^2/\mu^2)~ (f = e,\mu,\tau,u,d,c,s,b)$. For a process with $l$ external photons and $n$ EW couplings in the tree-level amplitude, if $l = n$, the full NLO EW correction is free of those unpleasant large logarithms because of the exact cancelation between the logarithms in vertex counterterms and in external photon wave-function counterterms; while if $l < n$, the uncanceled large logarithms can be absorbed into $n-l$ EW couplings by using running fine structure constant as input for these $n-l$ EW vertices. Therefore, we adopt the $G_\mu$-scheme for all the EW couplings of the $pp \rightarrow ZZ + {\rm jet} + X$ process in our calculation. In the $G_\mu$-scheme, the fine structure constant is chosen as
\begin{eqnarray}
\alpha_{G_\mu} = \frac{\sqrt{2}}{\pi}G_{\mu}M_W^2\sin^2\theta_W
\end{eqnarray}
to absorb those large logarithmic corrections in $\delta{Z^{\alpha(0)}_e}$. Correspondingly, the electric charge renormalization constant in the $G_\mu$-scheme should be modified as
\begin{eqnarray}
\delta Z_e^{G_\mu} = \delta Z_e^{\alpha(0)} -\frac{1}{2}\Delta r,
\end{eqnarray}
where $\Delta r$ is provided in Ref.\cite{delta} by considering the one-loop EW corrections to the muon decay. This subtraction term is introduced to avoid double counting in NLO EW calculation.

\subsection{Virtual EW corrections}
\par
\begin{figure}[htbp]
  \begin{center}
     \includegraphics[scale=0.85]{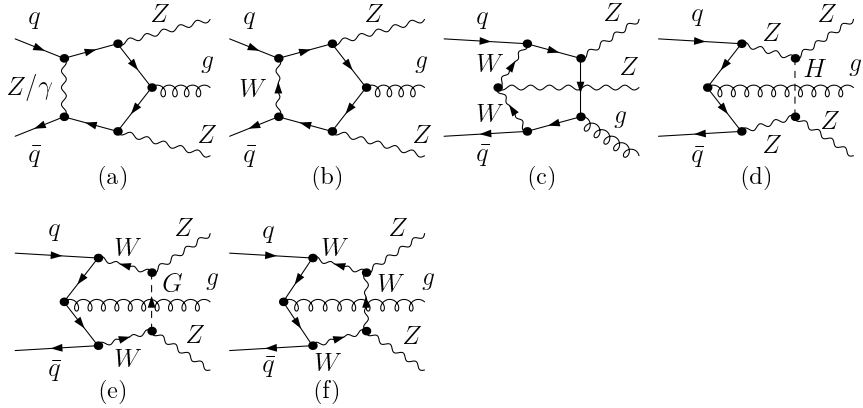}
    \caption{ Representative EW pentagon diagrams for the partonic process $q\bar{q}\rightarrow ZZg$.}
   \label{fig3}
  \end{center}
\end{figure}

The ${\cal O}(\alpha^3 \alpha_s)$ correction to the parent process $pp \rightarrow ZZ + {\rm jet} + X$ includes two parts: (1) NLO EW corrections to the quark-antiquark annihilation and quark-gluon fusion partonic channels (\ref{qqbar+qg}), and (2) NLO QCD corrections to the photon-induced subprocesses (\ref{gamma-induced}). The virtual corrections are induced by the related self-energy, vertex, box, and pentagon graphs. In Fig.\ref{fig3} we depict the representative EW pentagon diagrams for the partonic process $q\bar{q} \rightarrow ZZ g$. The UV divergences can be canceled exactly after performing the renormalization procedure, and the mass singularities are also removed since the $G_{\mu}$-scheme is adopted in the electric charge renormalization. The IR divergences originating from virtual photon exchange in loops can be canceled after adding the real photon emission corrections and the EW counterterms of parton distribution functions (PDFs). Then the final results are UV- and IR-finite.

\par
We follow the approach proposed by Denner and Dittmaier \cite{Denner} to decompose 5-point integrals into 4-point integrals. All tensor integrals can be reduced to scalar integrals recursively by using the Passarino-Veltman algorithm \cite{PV}. In the numerical calculation of 4-point integrals, we may confront a serious unstable problem induced by small Gram determinant. In order to solve this unstable problem, we add the quadruple precision arithmetic option in {\sc LoopTools}-2.8 \cite{looptools} by using the segmentation method analogous to that in Refs.\cite{detG1,detG2}. After the refinement above, the program can switch to the quadruple precision arithmetic in the region of ${\rm det} G_3/(2 k_{{\rm max}}^2)^3 < 10^{-5}$ flexibly, where ${\rm det} G_3$ is the Gram determinant and $k_{\rm max}^2$ the maximum of the external four-momentum squared for a given 4-point integral. Finally, we successfully keep the numerical instability under control and consume relatively less computer CPU time.

\subsection{Real photon emission corrections}
\par
The real photon emission corrections to the partonic channels in Eq.(\ref{qqbar+qg}) are from the following subprocesses:
\begin{eqnarray}
\label{real-r-emissions}
q \bar{q} \rightarrow ZZg\gamma,~~~~~~~
qg \rightarrow ZZq\gamma,~~~~~~~
\bar{q} g \rightarrow ZZ\bar{q}\gamma.
\end{eqnarray}
The soft and collinear IR divergences in real photon emission corrections are canceled exactly and partially with those from loop diagrams respectively, and the remaining collinear IR singularities are absorbed by the EW counterterms of quark PDFs. The quark PDF EW counterterm $\delta \Phi_{q|P}^{\rm{EW}}$ contains two parts: the collinear photon emission part $\delta \Phi_{q|P}^{\rm{EW},(\gamma)}$ and the collinear light-quark emission part $\delta \Phi_{q|P}^{{\rm EW},(q)}$. In the dimensional regularization and DIS factorization scheme, these two collinear parts are expressed as
\begin{eqnarray}
   \delta \Phi_{q|P}^{{\rm EW},(\gamma)}(x,\mu_f,\mu_r) =
   \frac{Q_q^2\alpha}{2 \pi}
   \int_x^1 \frac{dz}{z}
   \Phi_{q|P}(x/z,\mu_f)
   \biggl\{
   \frac{1}{\epsilon}
                      \frac{\Gamma(1 - \epsilon)}{\Gamma(1 - 2 \epsilon)}
                      \Big( \frac{4 \pi \mu_r^2}{\mu_f^2} \Big)^{\epsilon}
   \left[ P_{qq}(z) \right]_+
 - C_{qq}^{{\rm DIS}}(z)
   \biggr\}, \nonumber \\
   \delta \Phi_{q|P}^{{\rm EW},(q)}(x,\mu_f,\mu_r) =
   \frac{3Q_q^2\alpha}{2 \pi}
   \int_x^1 \frac{dz}{z}
   \Phi_{\gamma|P}(x/z,\mu_f)
   \biggl\{
   \frac{1}{\epsilon}
                      \frac{\Gamma(1 - \epsilon)}{\Gamma(1 - 2 \epsilon)}
                      \Big( \frac{4 \pi \mu_r^2}{\mu_f^2} \Big)^{\epsilon}
   P_{q\gamma}(z)
 - C_{q\gamma}^{{\rm DIS}}(z)
   \biggr\},~~
\end{eqnarray}
where $Q_q$ is the electric charge of quark $q$. The splitting functions $P_{ij}(z)$ are written as
\begin{eqnarray}
P_{qq}=\frac{1+z^2}{1-z}, ~~~~~~~
P_{q\gamma}(z)=z^2+(1-z)^2,
\end{eqnarray}
and the DIS subtraction functions $C_{ij}^{{\rm DIS}}(z)$ are given by \cite{detG1,DIS}
\begin{eqnarray}
&& C_{qq}^{{\rm DIS}}(z)
   =
   \biggl[
   P_{qq}(z) \Big( \ln \frac{1-z}{z} - \frac{3}{4} \Big) + \frac{9 + 5 z}{4} \biggr]_+ , \nonumber \\
&& C_{q\gamma}^{{\rm DIS}}(z)
   =
   P_{q\gamma}(z)
   \ln \frac{1-z}{z} - 8z^2 + 8z - 1.~~~~~~~~~~~~~~
\end{eqnarray}
The $\left[ \ldots \right]_+$ prescription is understood as
\begin{eqnarray}
\int_0^1 dz \left[ g(z) \right]_+ f(z) = \int_0^1 dz \, g(z) \left[ f(z) - f(1) \right].
\end{eqnarray}

\par
In calculating real photon emission partonic processes, we employ the two cutoff phase space slicing (TCPSS) method \cite{two-cutoff} to isolate the soft and collinear IR singularities. By introducing two arbitrary small cutoffs $\delta_s$ and $\delta_c$, the phase space of a real photon emission process is decomposed into soft ($E_{\gamma} \leq \delta_{s} \sqrt{\hat{s}}/2$), hard collinear ($E_{\gamma} > \delta_{s} \sqrt{\hat{s}}/2$, $\min\{\hat{s}_{\gamma f}\} \leq \delta_{c}\hat{s}$) and hard noncollinear ($E_{\gamma} > \delta_{s} \sqrt{\hat{s}}/2$, $\min\{\hat{s}_{\gamma f}\} > \delta_{c}\hat{s}$) regions, where $\sqrt{\hat{s}}$ is the partonic center-of-mass colliding energy, $\hat{s}_{ij} = (p_i+p_j)^2$ and $f$ runs over the charged fermions in initial and final states. The soft and collinear IR singularities are located in the soft and hard collinear regions, respectively, while the phase space integration over the hard noncollinear region is IR-finite. The cutoff independence of the real photon emission corrections have been checked numerically in the range of $10^{-6} < \delta_s < 10^{-3}$ with $\delta_{c} = \delta_{s}/50$.

\subsection{Event identification and selection}
\label{EISS}
\par
For the inclusive $ZZ + {\rm jet}$ production, there exist $ZZ + {\rm jet} + {\rm jet}$ and $ZZ + {\rm jet} + \gamma$ four-particle events originating from the real gluon, light-quark and photon emissions at ${\cal O}(\alpha^2 \alpha_s^2)$ and ${\cal O}(\alpha^3 \alpha_s)$. The topologies of these real emission subprocesses are expressed as
\begin{eqnarray}
\label{topology-real}
0 \rightarrow ZZ q\bar{q}gg,~~~~~~
0 \rightarrow ZZ q\bar{q} q^{\prime}\bar{q}^{\prime},~~~~~~
0 \rightarrow ZZ q\bar{q} g\gamma.
\end{eqnarray}
We apply the transverse momentum cut of
\begin{eqnarray}
\label{pTjet-cut}
p_{T,{\rm jet}} > p_{T,{\rm jet}}^{{\rm cut}}
\end{eqnarray}
on the leading jet of final state to ensure a detectable hard jet in the inclusive $ZZ + {\rm jet}$ production. This kinematic cut can also guarantee the IR safety at the LO. If the two tracks of jets (or jet and photon) of final state are sufficiently collinear, i.e., $R_{{\rm jet} {\rm jet}}~ ({\rm or}~ R_{\gamma {\rm jet}}) < R_0$, where $R_{ij} = \sqrt{(y_i - y_j)^2 + (\phi_i - \phi_j)^2}$ represents the separation of the two tracks on the rapidity-azimuthal-angle plane, we merge them into a single jet track. However, this naive track combination procedure is always accompanied with two problems for $0 \rightarrow ZZ q\bar{q} g\gamma$ if the final state is $ZZ+{\rm jet}+\gamma$. First, the $0 \rightarrow ZZ q\bar{q} g\gamma$ topology with $ZZ+{\rm jet}+\gamma$ final state, i.e., the partonic processes in Eq.(\ref{real-r-emissions}), can be treated as not only the real photon bremsstrahlung to the $ZZ + {\rm jet}$ production but also the real jet emission to the $ZZ + \gamma$ production. Therefore, the jet-photon-merged track might be regarded as a photon, not necessarily a jet. Second, for the $q\bar{q} \rightarrow ZZ g \gamma$ subprocess, if the gluon-photon-merged track is regarded as a jet, the energy fraction of the gluon inside the jet can be arbitrarily small even if the jet selection criterion (\ref{pTjet-cut}) is applied. This soft gluon induces an unexpected QCD soft IR singularity that cannot be canceled at the EW NLO.

\par
To solve these two problems we introduce an event selection criterion for the $pp \rightarrow ZZ + {\rm jet} + \gamma + X$ process in which $ZZ + {\rm jet}$ and $ZZ + \gamma$ events are properly defined. In the case that $R_{\gamma {\rm jet}} < R_0$, the jet and photon tracks are merged into a single track and the final state is a three-particle event. If $z_{\gamma} > z_{\gamma}^{{\rm cut}}$ where $z_{\gamma}$ is the energy fraction of the photon inside the merged track, this three-particle event is called as a $ZZ + \gamma$ event and rejected; otherwise, it is treated as a $ZZ + {\rm jet}$ event and kept \cite{frag1,frag2,frag3}. However, this event selection criterion leads to the uncanceled final-state QED collinear IR divergence from the $qg \rightarrow ZZq\gamma$ and $\bar{q}g \rightarrow ZZ\bar{q}\gamma$ partonic processes in the region of $z_{\gamma} \in (z_{\gamma}^{{\rm cut}}, 1]$. In analogy to the absorption of initial-state collinear IR singularities into PDFs, this remaining QED collinear IR divergence can be absorbed into the NLO definition of the quark-to-photon fragmentation function.

\par
At the EW NLO, the bare quark-to-photon fragmentation function in the $\overline{{\rm MS}}$ renormalization scheme can be written as \cite{LEP1}
\begin{eqnarray}
D^{{\rm bare}}_{q \rightarrow \gamma}(z_\gamma)
=
\frac{Q_q^2 \alpha}{2\pi} \frac{1}{\epsilon} \frac{1}{\Gamma(1-\epsilon)}
\Big( \frac{4 \pi \mu_r^2}{\mu_f^2} \Big)^{\epsilon}
P_{\gamma q}(z_{\gamma})
+
D_{q \rightarrow \gamma}(z_\gamma,\mu_f),
\end{eqnarray}
where the quark-to-photon splitting function $P_{\gamma q}(z_{\gamma})$ is given by
\begin{eqnarray}
P_{\gamma q}(z_{\gamma})
= \frac{1 + (1 - z_{\gamma})^2}{z_{\gamma}}.
\end{eqnarray}
The nonperturbative fragmentation function $D_{q \rightarrow \gamma}(z_\gamma,\mu_f)$ is experimentally feasible and has been measured at the LEP in $\gamma + {\rm jet}$ events. In this paper, we employ the parametrization of the nonperturbative fragmentation function used by the ALEPH collaboration \cite{LEP2}, i.e.,
\begin{eqnarray}
D_{q \rightarrow \gamma}(z_\gamma,\mu_f)
=
D_{q \rightarrow \gamma}^{{\rm ALEPH}}(z_\gamma,\mu_f)
\equiv
\frac{Q_q^2 \alpha}{2\pi}
\biggl( P_{\gamma q}(z_\gamma) \ln\frac{\mu^2_f}{(1-z_\gamma)^2 \mu^2_0} + C \biggr),
\end{eqnarray}
where $\mu_0 = 0.14~ {\rm GeV}$ and $C = -1 - \ln(M_Z^2/2\mu_0^2)= -13.26$ are obtained from one-parameter data fit.

\par
According to the event selection criterion described above, we should subtract the contribution of the $ZZ + \gamma$ events from the perturbatively well-defined inclusive cross section in which the photon energy fraction $z_{\gamma}$ ranges over $0 \leq z_\gamma \leq 1$. The subtraction term can be written as
\begin{eqnarray}
\label{sub-term}
d \sigma^{{\rm (sub)}}
=
d \sigma^{{\rm (pert)}}_{{\rm fin}}
+
\Big[ d \sigma^{{\rm (pert)}}_{{\rm sing}} + d \sigma^{{\rm (frag)}} \Big],
\end{eqnarray}
where $d \sigma^{{\rm (pert)}}$ and $d \sigma^{{\rm (frag)}}$ correspond to the perturbative radiation and nonperturbative production of a photon over the region of $z_{\gamma} \in (z_{\gamma}^{{\rm cut}}, 1]$, and the subscripts ``fin" and ``sing" denote the collinear-safe and -singular parts respectively. By employing the TCPSS method, the two terms of the right-hand side of Eq.(\ref{sub-term}) can be expressed as
\begin{eqnarray}
\label{pert+frag}
&&~~~~~~~~~~~~~~~~ d \sigma^{{\rm (pert)}}_{{\rm fin}}
=
\sum_{q = u,d,c,s,b}
\int^1_{z_{{\rm cut}}}
\biggl[
d \sigma^{{\rm LO}}(pp \rightarrow q\bar{q} \rightarrow ZZ+ {\rm jet} + \gamma)\Big|_{R_{\gamma {\rm jet}} < R_0} \nonumber \\
&&~~~~~~~~~~~~~~~~~~~~~~~~~~~~ ~~~~~~~~~~~~ +\,
d \sigma^{{\rm LO}}(pp \rightarrow qg, \bar{q}g \rightarrow ZZ+ {\rm jet} + \gamma)\Big|_{R_{\gamma {\rm jet}} < R_0,~ \hat{s}_{\gamma {\rm jet}} > \delta_{c}\hat{s}} \biggr],~~~~~~~~
\nonumber \\
&& \Big[ d \sigma^{{\rm (pert)}}_{{\rm sing}} + d \sigma^{{\rm (frag)}} \Big]
=
\sum_{q = u,d,c,s,b}
d \sigma^{{\rm LO}}(pp \rightarrow qg, \bar{q}g \rightarrow ZZ + {\rm jet})
\int^1_{z_{{\rm cut}}} dz_\gamma {\cal D}_{q \rightarrow \gamma}(z_\gamma),
\end{eqnarray}
where the (collinear-safe) effective quark-to-photon fragmentation function ${\cal D}_{q \rightarrow \gamma}(z_{\gamma})$ is defined as \cite{LEP1}
\begin{eqnarray}
\label{D-eff}
{\cal D}_{q \rightarrow \gamma}(z_\gamma)
=
-\frac{Q_q^2 \alpha}{2\pi} \frac{1}{\epsilon} \frac{1}{\Gamma(1-\epsilon)}
\Big( \frac{4 \pi \mu_r^2}{\delta_c \hat{s}} \Big)^{\epsilon}
[ z_\gamma (1 - z_\gamma) ]^{-\epsilon}
[ P_{\gamma q}(z_\gamma) - \epsilon z_{\gamma} ]
+
D^{{\rm bare}}_{q \rightarrow \gamma}(z_\gamma).
\end{eqnarray}
As shown in Eqs.(\ref{sub-term}), (\ref{pert+frag}) and (\ref{D-eff}), the subtraction term $d \sigma^{{\rm (sub)}}$ is collinear-safe, therefore an (UV- and) IR-finite NLO QCD+EW corrected cross section for the $pp \rightarrow ZZ + {\rm jet} + X$ is obtained after applying the $ZZ + {\rm jet}$ event selection criterion.

\vskip 5mm
\section{Numerical results and discussion}
\subsection{Input parameters and setup}
\par
The relevant SM input parameters are \cite{PDG}
\begin{eqnarray}
&&
M_W = 80.385~ {\rm GeV},~~~~~
M_Z = 91.1876~ {\rm GeV},~~~~~
M_H = 125.7~ {\rm GeV}, \nonumber \\
&&
m_e = 0.510998928~ {\rm MeV},~~~~
m_{\mu} = 105.6583715~ {\rm MeV},~~~~
m_{\tau} = 1.77682~ {\rm GeV},~~~~~~~~~~~~ \nonumber \\
&&
m_t = 173.21~ {\rm GeV},~~~~~
G_{\mu} = 1.1663787 \times 10^{-5}~ {\rm GeV}^{-2},~~~~~
\alpha_s(M_Z) = 0.119.~~~~~~~~~~~
\end{eqnarray}
If the Cabibbo-Kobayashi-Maskawa (CKM) matrix $V_{{\rm CKM}}$ is $2 \oplus 1$ block-diagonal, i.e., only the quark mixing between the first two generations is taken into account, the LO and NLO corrected cross sections for the $ZZ + {\rm jet}$ production are independent of the CKM matrix elements because all the related topologies\footnote{The topologies related to the $ZZ + {\rm jet}$ production at QCD+EW NLO are $0 \rightarrow ZZq\bar{q}g$ and those in Eq.(\ref{topology-real}).} contain no charged-current quark chain. Therefore, we set $V_{{\rm CKM}} = \mathbf{1}_{3 \times 3}$ in numerical calculation.

\par
We employ the NLO NNPDF2.3QED PDFs \cite{NNPDF} with $\overline{{\rm MS}}$ and DIS factorization schemes in the NLO QCD and EW calculations, respectively. The strong coupling constant $\alpha_s$ is renormalized in the $\overline{{\rm MS}}$ scheme with five active flavors. The NLO QCD and EW corrections are expressed as
\begin{eqnarray}
\Delta \sigma^{\textrm{NLO QCD}}
=
\sigma_{\textrm{virt}}^{\alpha_s}
+
\sigma_{\textrm{real}}^{\alpha_s}
+
\sigma_{\textrm{pdf}}^{\alpha_s}
+
\big( \sigma^0 - \sigma^{\textrm{LO}} \big), \nonumber \\
\Delta \sigma^{\textrm{NLO EW}}
=
\sigma_{\textrm{virt}}^{\alpha}
+
\sigma_{\textrm{real}}^{\alpha}
+
\sigma_{\textrm{pdf}}^{\alpha}
-
\sigma^{\textrm{(sub)}},~~~~~~~\,
\end{eqnarray}
where $\sigma_{\textrm{virt}}^{\alpha_s,\alpha}$, $\sigma_{\textrm{real}}^{\alpha_s,\alpha}$ and $\sigma_{\textrm{pdf}}^{\alpha_s,\alpha}$ are the virtual, real and PDF-counterterm corrections at ${\cal O}(\alpha^2 \alpha_s^2)$ and ${\cal O}(\alpha^3 \alpha_s)$, respectively, and $\sigma^{\textrm{LO}}$ and $\sigma^{0}$ are the LO cross sections calculated with the LO and NLO NNPDF2.3QED PDFs, separately. As discussed in Section \ref{general-description}, we also include the contributions from the photon-induced subprocesses (\ref{gamma-induced}) up to ${\cal O}(\alpha^3 \alpha_s)$, i.e.,
\begin{eqnarray}
\sigma_{\gamma\textrm{-ind}} = \sigma_{\gamma\textrm{-ind}}^{0} + \Delta \sigma_{\gamma\textrm{-ind}}^{\textrm{NLO QCD}},
\end{eqnarray}
where $\sigma_{\gamma\textrm{-ind}}^{0}$ and $\Delta \sigma_{\gamma\textrm{-ind}}^{\textrm{NLO QCD}}$ are Born and NLO QCD photon-induced contributions, respectively, both calculated with the NLO
NNPDF2.3QED PDFs. Then the relative QCD, EW and photon-induced corrections are given by
\begin{eqnarray}
\delta_{\textrm{QCD}}
=
\frac{\Delta \sigma^{\textrm{NLO QCD}}}{\sigma^{\textrm{LO}}},~~~~~~~~
\delta_{\textrm{EW}}
=
\frac{\Delta \sigma^{\textrm{NLO EW}}}{\sigma^{0}},~~~~~~~~
\delta_{\gamma\textrm{-ind}}
=
\frac{\sigma_{\gamma\textrm{-ind}}}{\sigma^{\textrm{LO}}}.
\end{eqnarray}
To obtain the full NLO corrected cross section, we combine the QCD and EW corrections by using the naive product of the relative corrections and add the photon-induced contributions linearly \cite{Multiply-scheme}, i.e.,
\begin{eqnarray}
\sigma^{\textrm{NLO}}
=
\sigma^{\textrm{LO}}
\Big[
( 1 + \delta_{\textrm{QCD}} )
( 1+ \delta_{\textrm{EW}} )
+
\delta_{\gamma\textrm{-ind}}
\Big]
=
\sigma^{{\textrm{LO}}}
( 1 + \delta_{\textrm{NLO}} ).
\end{eqnarray}

\par
The parameters for $ZZ + {\rm jet}$ event identification and selection are fixed as
\begin{eqnarray}
R_0 = 0.5,~~~~~~~
z_{{\rm cut}} = 0.7,~~~~~~~
p_{T,{\rm jet}}^{{\rm cut}} = 50~ {\rm GeV}.
\end{eqnarray}
The factorization and renormalization scales are set to be equal $(\mu_f = \mu_r = \mu)$ for simplicity, and the central scale value is chosen as
\begin{eqnarray}
\mu_0 = H_T/2 = \sum m_{T}/2,
\end{eqnarray}
where $m_T = \sqrt{m^2 + \vec{p}_T^{\,2}}$ is the transverse mass of a particle and the summation is taken over all the final particles for the process $pp \rightarrow ZZ + {\rm jet} + X$. In the following numerical calculation, we take $\mu = \mu_0$ by default unless otherwise stated. Compared to the fixed scale choice used in Ref.\cite{zzj-qcd}, this dynamic factorization/renormalization scale would be better to capture the information of dynamics than the fixed one.

\subsection{Integrated cross sections}
\par
In Table \ref{table1} we list the LO, NLO QCD+EW corrected (including photon-induced contributions) integrated cross sections and the corresponding relative corrections ($\delta_{\textrm{QCD}}$, $\delta_{\textrm{EW}}$, $\delta_{\gamma\textrm{-ind}}$ and $\delta_{\textrm{NLO}}$) for the $ZZ + {\rm jet}$ production at $\sqrt{S} = 13,~14,~33$ and $100~{\rm TeV}$ hadron colliders separately. We can see from the table that the LO cross section is enhanced by the NLO QCD correction while suppressed by the NLO EW correction. Although the LO cross section increases notably with the increment of the proton-proton colliding energy, both the relative QCD and EW corrections are insensitive to the colliding energy. The NLO EW correction is about one order of magnitude smaller than the NLO QCD correction but quantitatively not negligible, while the photon-induced correction is very small compared to both the NLO QCD and NLO EW corrections.
\begin{table}[h]
\begin{center}
\begin{tabular}{ccccccc}
\hline
\hline
\\ [-1.7ex]
$\sqrt{S}$~[TeV] & $\sigma^\textrm{LO}$~[pb] & $\sigma^\textrm{NLO}$~[pb] & $\delta_\textrm{QCD}$~[\%] & $\delta_\textrm{EW}$~[\%] & $\delta_{\gamma\textrm{-ind}}$~[\%] & $\delta_\textrm{NLO}$~[\%] \\
\\ [-1.7ex]
\hline
\\ [-1.7ex]
$13$ & $1.8709(1)$ & $2.708(4)$  & $52.6$ & $-5.22$ & $0.13$ & $44.76$ \\
\\ [-1.7ex]
$14$ & $2.1348(3)$  & $3.087(5)$ & $52.6$ & $-5.32$ & $0.13$ & $44.61$ \\
\\ [-1.7ex]
$33$ & $8.6670(8)$  & $12.63(2)$ & $54.4$ & $-5.66$ & $0.10$ & $45.76$ \\
\\ [-1.7ex]
$100$  & $41.916(5)$ & $60.45(8)$ & $53.5$ & $-6.10$ & $0.07$ & $44.21$ \\
\\ [-1.7ex]
\hline
\hline
\end{tabular}
\caption{The LO, NLO QCD+EW corrected integrated cross sections and the corresponding relative corrections for the $ZZ + {\rm jet}$ production at $\sqrt{S} = 13,~14,~33$ and $100~{\rm TeV}$ proton-proton colliders.}
\label{table1}
\end{center}
\end{table}

\par
In Table \ref{table2} we demonstrate the $p_{T,{\rm jet}}^{{\rm cut}}$ dependence of the LO, NLO QCD+EW corrected integrated cross sections and the corresponding relative corrections for the $ZZ + {\rm jet}$ production at the $14~ {\rm TeV}$ LHC. The table shows that both the QCD and EW relative corrections are sensitive to the transverse momentum cut on the hardest jet. With the increment of $p_{T,{\rm jet}}^{{\rm cut}}$, the LO cross section decreases quickly since the transverse momentum cut of $p_{T,{\rm jet}} > p_{T,{\rm jet}}^{{\rm cut}}$ is imposed on the hardest jet to select the $ZZ + {\rm jet}$ events, while the relative QCD correction increases due to the experimentally unresolved real jet radiation at the QCD NLO. Contrary to the QCD correction, the relative EW correction decreases with the increment of $p_{T,{\rm jet}}^{{\rm cut}}$, because the real photon emission would soften the final jet, and moreover, the final state is not $ZZ + {\rm jet}$ event if the photon is unresolved and sufficiently energetic.
\begin{table}[h]
\begin{center}
\begin{tabular}{ccccccc}
\hline
\hline
\\ [-1.7ex]
$p_{T,{\rm jet}}^{{\rm cut}}$~[GeV] & $\sigma^\textrm{LO}$~[pb] & $\sigma^\textrm{NLO}$~[pb] & $\delta_\textrm{QCD}$~[\%] & $\delta_\textrm{EW}$~[\%] & $\delta_{\gamma\textrm{-ind}}$~[\%] & $\delta_\textrm{NLO}$~[\%] \\
\\ [-1.7ex]
\hline
\\ [-1.7ex]
$20$ & $5.2701(6)$ & $7.146(9)$ & $42.0$ & $-4.59$ & $0.11$ & $35.59$  \\
\\ [-1.7ex]
$50$ & $2.1348(3)$  & $3.087(5)$ & $52.6$ & $-5.32$ & $0.13$ & $44.61$ \\
\\ [-1.7ex]
$100$  & $0.76528(7)$ & $1.176(2)$ & $65.2$ & $-7.04$ & $0.16$ & $53.73$ \\
\\ [-1.7ex]
$200$  & $0.16125(2)$ & $0.2759(4)$ & $91.8$ & $-10.91$ & $0.20$ & $71.07$ \\
\\ [-1.7ex]
\hline  \hline
\end{tabular}
\caption{The LO, NLO QCD+EW corrected integrated cross sections and the corresponding relative corrections for the $ZZ + {\rm jet}$ production at the $14~{\rm TeV}$ LHC by taking $p_{T,{\rm jet}}^{{\rm cut}} = 20,~ 50,~ 100$ and $200~ {\rm GeV}$.}
\label{table2}
\end{center}
\end{table}

\par
To estimate the theoretical uncertainty from the factorization and renormalization scales, we define the upper and lower relative scale uncertainties as
\begin{eqnarray}
\eta_{+,\,-}
=
\textrm{max,\,min}\left\{ \frac{\sigma(\mu_f,\mu_r)}{\sigma(\mu_0,\mu_0)}\, \biggl|\, \mu_f, \mu_r \in \{ \mu_0/2,~ 2 \mu_0 \} \right\} - 1.
\end{eqnarray}
The LO and NLO QCD+EW corrected integrated cross sections combined with the scale uncertainties for the $ZZ + {\rm jet}$ production at the $14~{\rm TeV}$ LHC are given as
\begin{eqnarray}
\sigma^{\textrm{LO}}= 2.1348^{+9.8\%}_{-8.5\%}~\textrm{pb},~~~~~~~~\sigma^{\textrm{NLO}}=3.087^{+5.5\%}_{-4.7\%}~\textrm{pb}.
\end{eqnarray}
It shows that the NLO QCD+EW correction can improve the accuracy of the integrated cross section by reducing the scale uncertainty.

\subsection{Kinematic distributions}
\par
In this subsection we provide some kinematic distributions of final particles before and after the $Z$-boson leptonic decays at the $14~{\rm TeV}$ LHC.

\subsubsection{Distributions for $pp \rightarrow ZZ + {\rm jet} + X$}
\par
\begin{figure}[htbp]
\includegraphics[width=10cm]{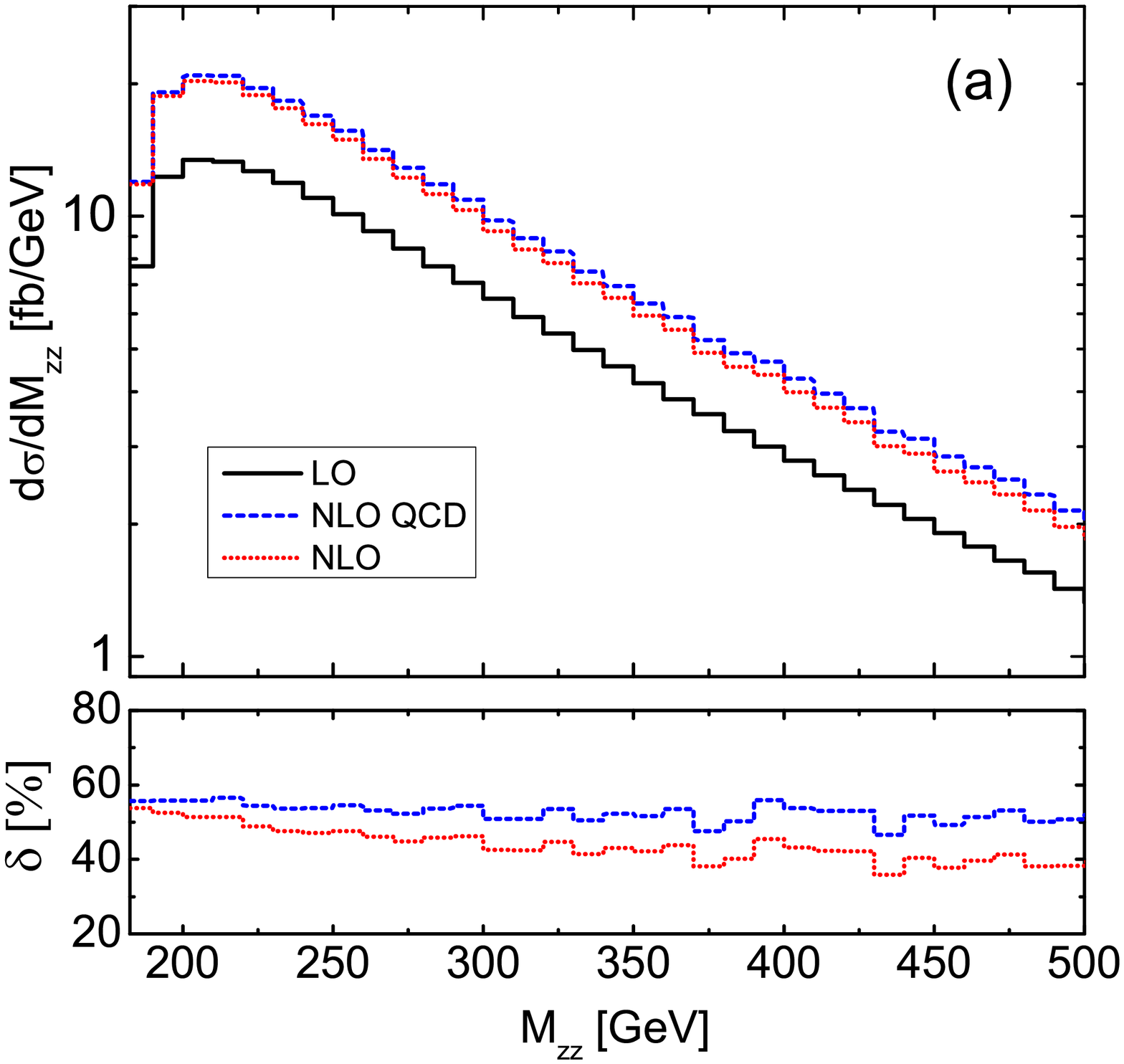}
\hspace{-2cm}
\includegraphics[width=10cm]{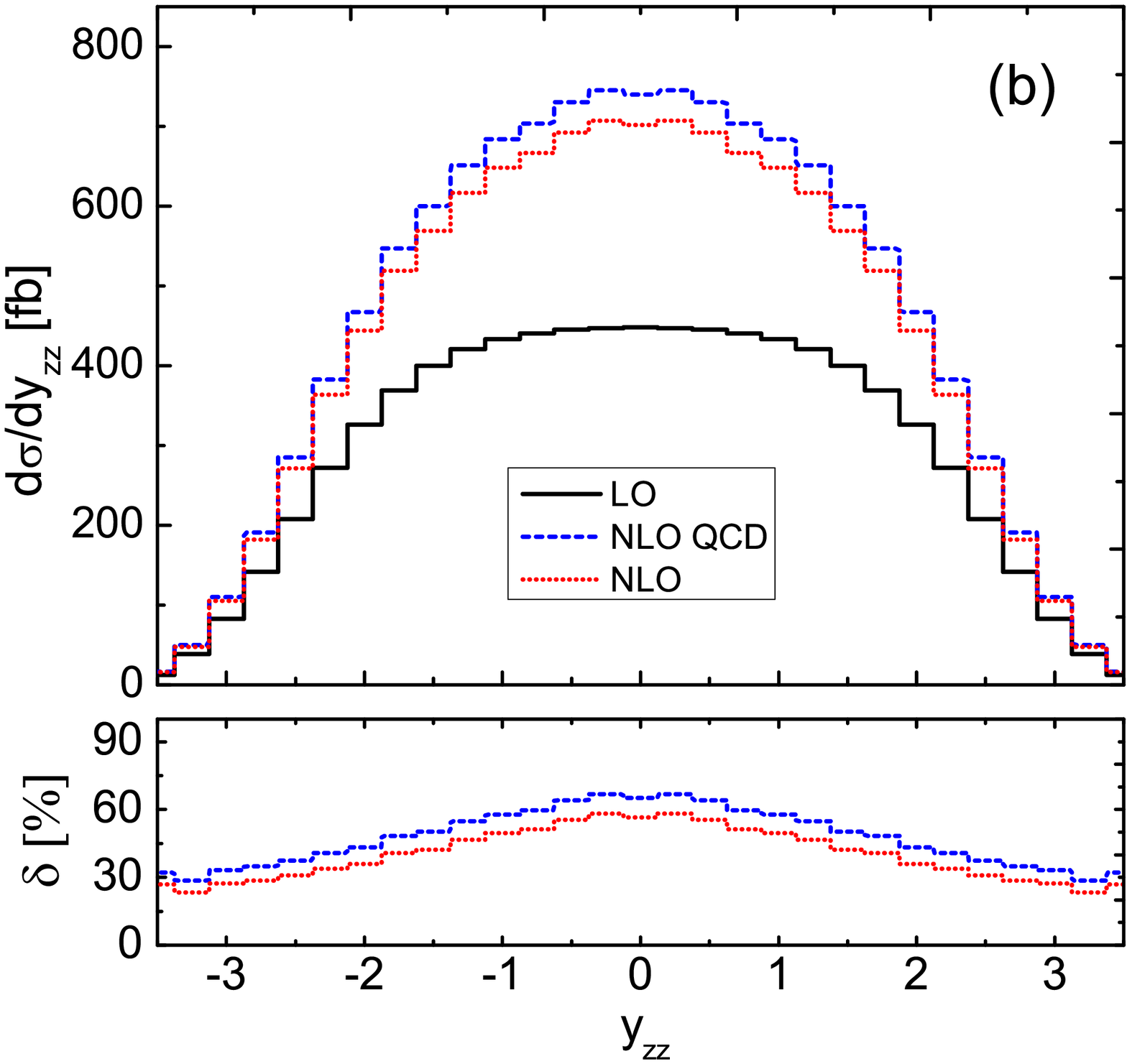}
\caption{  The LO, NLO QCD and NLO QCD+EW corrected (a) invariant mass and (b) rapidity distributions of $Z$-boson pair for $pp \rightarrow ZZ + {\rm jet} + X$ at the $14~{\rm TeV}$ LHC. }
\label{fig-zz}
\end{figure}

In Fig.\ref{fig-zz}(a) we present the LO, NLO QCD, NLO QCD+EW corrected $Z$-pair invariant mass distributions ($\frac{d\sigma^{\textrm{LO}}}{dM_{ZZ}}$, $\frac{d\sigma^{\textrm{NLO QCD}}}{dM_{ZZ}}$, $\frac{d\sigma^{\textrm{NLO}}}{dM_{ZZ}}$) and the corresponding relative corrections for $pp \rightarrow ZZ + {\rm jet} + X$. From the figure we see that the NLO QCD and EW corrections do not distort the line shape of the LO $M_{ZZ}$ distribution. The NLO QCD correction enhances the LO $M_{ZZ}$ distribution significantly, while the NLO EW correction is small compared to the NLO QCD correction and slightly suppresses the LO $M_{ZZ}$ distribution. Both the LO and NLO corrected $M_{ZZ}$ distributions reach their maxima in the vicinity of $M_{ZZ} \sim 200 ~{\rm GeV}$, and then decrease rapidly with the increment of $M_{ZZ}$. In the plotted $M_{ZZ}$ region, the relative QCD correction is stable, while the relative EW correction decreases from $-1.24\%$ to $-8.30\%$ with the increment of $M_{ZZ}$. The NLO EW correction becomes relatively significant in high $M_{ZZ}$ region due to the large EW Sudakov logarithms. The full NLO (QCD+EW) relative correction decreases from $53.7\%$ to $38.3\%$ as the increment of $M_{ZZ}$ from its threshold to $500~ {\rm GeV}$.

\par
The LO, NLO QCD and NLO QCD+EW corrected rapidity distributions of $Z$-boson pair ($\frac{d\sigma^{\textrm{LO}}}{dy_{ZZ}}$, $\frac{d\sigma^{\textrm{NLO QCD}}}{dy_{ZZ}}$ and $\frac{d\sigma^{\textrm{NLO}}}{dy_{ZZ}}$) are depicted in Fig.\ref{fig-zz}(b), and the corresponding relative corrections are plotted in the lower panel. We see clearly that the relative QCD correction is positive and decreases with the increment of $|y_{ZZ}|$, while the relative EW correction is insensitive to $y_{ZZ}$ and supresses the LO $y_{ZZ}$ distribution a little bit in the whole plotted $y_{ZZ}$ region. At $y_{ZZ} = 0$ and $\pm 3$, the relative QCD corrections are $65.1\%$ and $33.1\%$, while the relative EW corrections are $-5.18\%$ and $-4.30\%$, respectively. Consequently, we get the full NLO relative correction as $56.5\%$ at $y_{ZZ}=0$ and $27.4\%$ at $|y_{ZZ}|=3$.

\par
\begin{figure}[htbp]
\includegraphics[width=10cm]{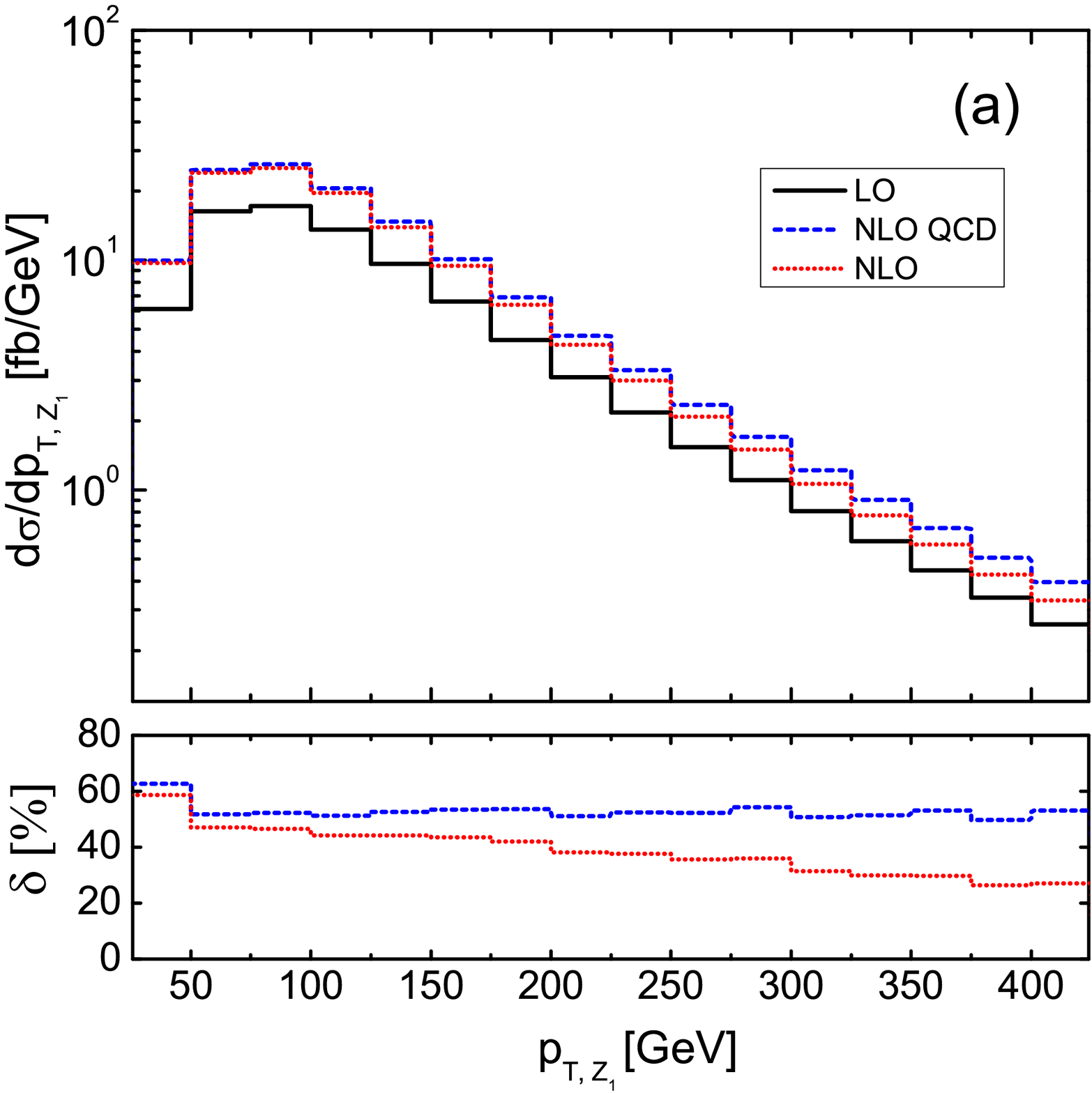}
\hspace{-2cm}
\includegraphics[width=10cm]{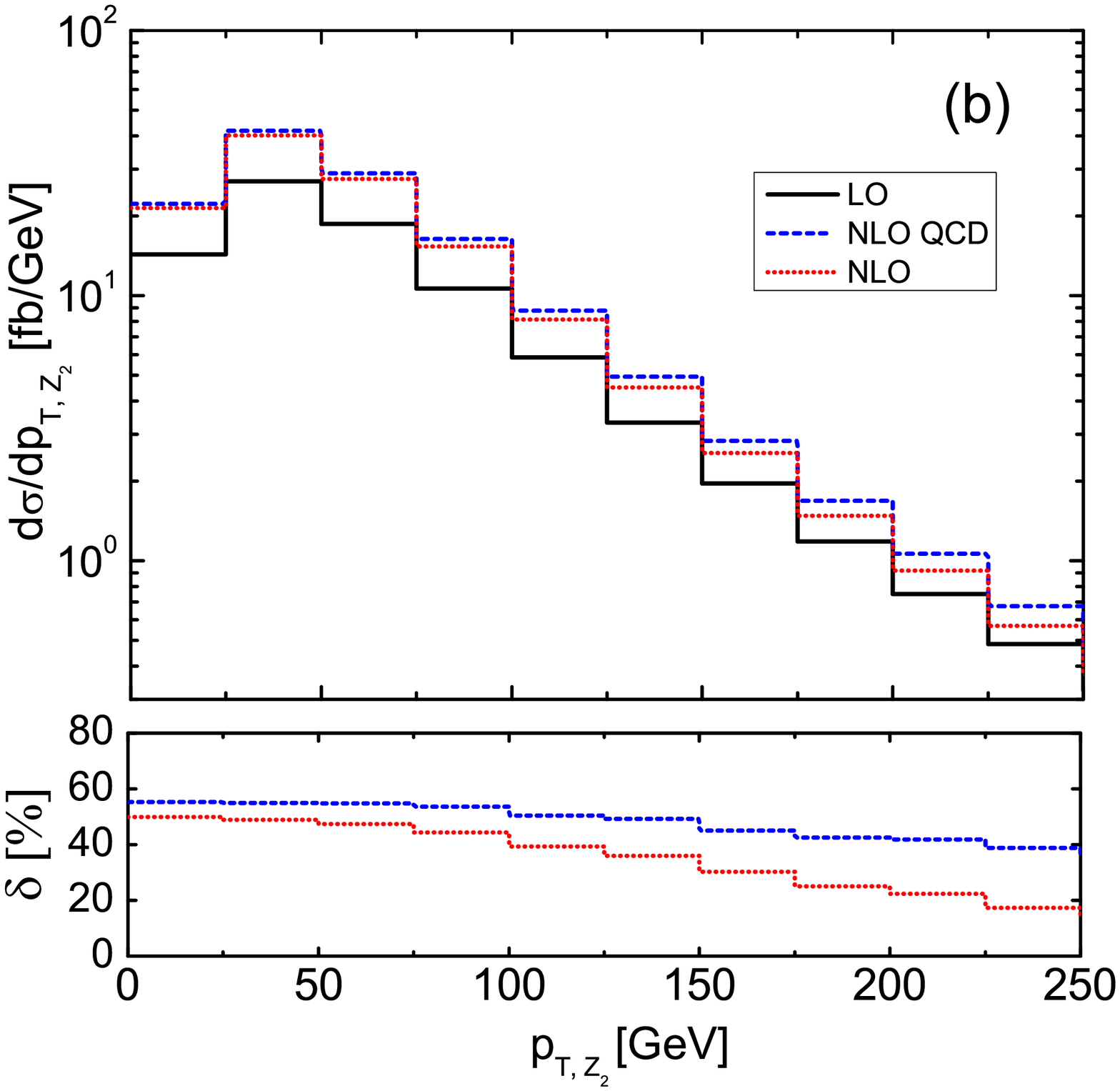}
\caption{ The LO, NLO QCD and NLO QCD+EW corrected transverse momentum distributions of (a) leading and (b) second $Z$-bosons for $pp \rightarrow ZZ + {\rm jet} + X$ at the $14~{\rm TeV}$ LHC. }
\label{fig-z}
\end{figure}

Among the two final $Z$-bosons, the leading $Z$-boson $Z_1$ and the second $Z$-boson $Z_2$ are defined as
\begin{eqnarray}
p_{T,Z_1} > p_{T,Z_2}.
\end{eqnarray}
Their transverse momentum distributions and the corresponding relative corrections are shown in Figs.\ref{fig-z}(a) and (b) separately. We see that both the LO and NLO corrected transverse momentum distributions peak at $p_{T,Z_1} \sim 75~ {\rm GeV}$ and $p_{T,Z_2} \sim 25~ {\rm GeV}$ for the leading and second $Z$-bosons, respectively. The relative QCD correction is steady at about $50\%$ in the plotted $p_{T,Z_1}$ region for the leading $Z$-boson, while decreases from $55.2\%$ to $38.8\%$ as $p_{T,Z_2}$ increases from $0$ to $250~ {\rm GeV}$ for the second $Z$-boson. However, the relative EW corrections to both $p_{T,Z_1}$ and $p_{T,Z_2}$ distributions decrease, from $-2.46\%$ to $-17.0\%$ and from $-3.45\%$ to $-15.5\%$ respectively, with the increment of $p_{T,Z_1}$ and $p_{T,Z_2}$ in their plotted regions. In analogy to the $M_{ZZ}$ distribution, the large relative EW correction in high $p_T$ region is due to the EW Sudakov effect. Consequently, the NLO QCD+EW correction enhances the LO transverse momentum distributions of the leading and second $Z$-bosons, and the corresponding NLO relative corrections decrease from $58.7\%$ to $27.1\%$ and from $49.9\%$ to $17.3\%$ respectively as the increment of $p_{T,Z_1} \in [25, 425]~ {\rm GeV}$ and $p_{T,Z_2} \in [0, 250]~{\rm GeV}$.

\subsubsection{Distributions for $pp \rightarrow ZZ + {\rm jet} \rightarrow 4 \ell + {\rm jet} + X$}
\par
Now we turn to the $ZZ + {\rm jet}$ production with subsequent $Z$-boson leptonic decays, i.e., $pp \rightarrow ZZ + {\rm jet} \rightarrow \ell^{+} \ell^{-} \ell^{\prime +} \ell^{\prime -} + {\rm jet} + X~ (\ell, \ell^{\prime} = e, \mu, \tau)$, at the $14~ {\rm TeV}$ LHC. For each same-sign lepton pair in the final state ($\ell^+ \ell^{\prime +}$ or $\ell^- \ell^{\prime -}$), the lepton with larger transverse momentum is called the leading lepton $\ell_1$ and the other the second lepton $\ell_2$. In the following we provide and discuss the distributions of the transverse momenta and azimuthal-angle separation of the two negatively charged leptons, i.e., $\frac{d\sigma}{dp_{T,\ell_1^-}}$, $\frac{d\sigma}{dp_{T,\ell_2^-}}$ and $\frac{d\sigma}{d\phi_{\ell_1^-\ell_2^-}}$. In order to taken into account the off-shell contribution and spin correlation from the $Z$-boson leptonic decays, we transform the differential cross sections into Les Houches event files \cite{LHE1,LHE2} and use MadSpin method \cite{Madspin1, Madspin2} to obtain events after the $Z$-boson decays.

\par
\begin{figure}[htbp]
\includegraphics[width=10cm]{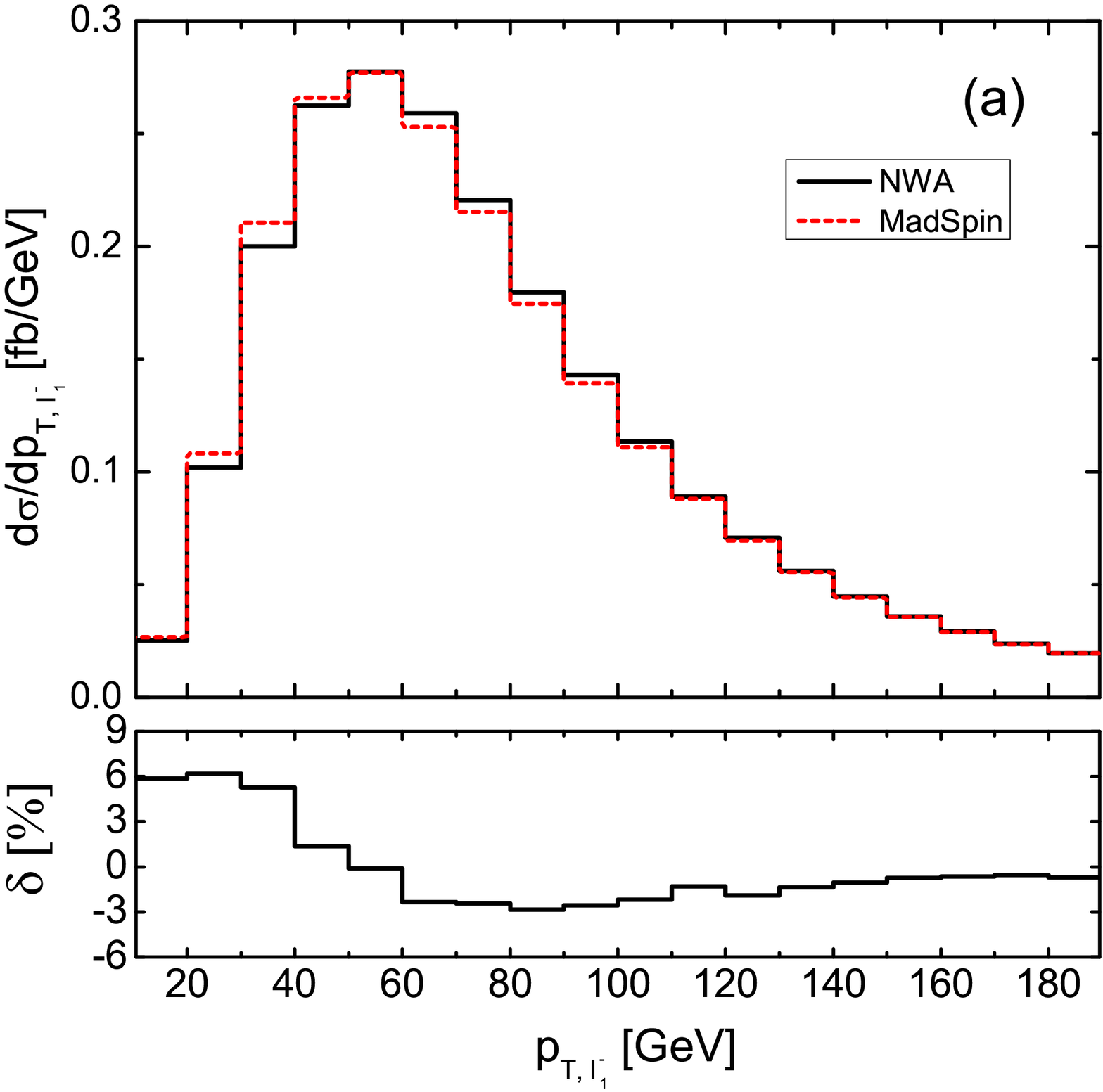}
\hspace{-2cm}
\includegraphics[width=10cm]{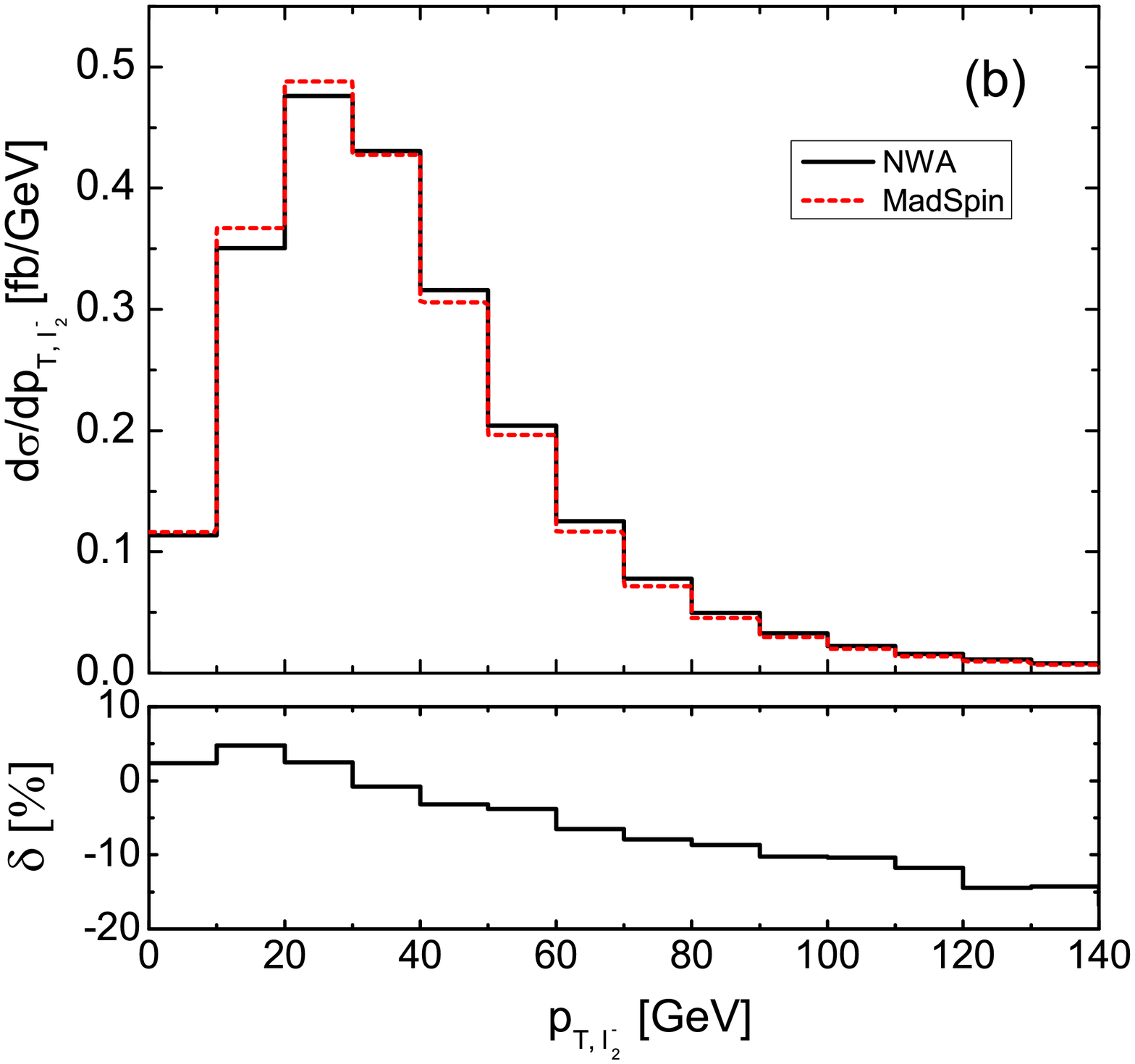}
\caption{ The LO transverse distributions of (a) leading and (b) second negatively charged leptons for $pp \rightarrow ZZ + {\rm jet} \rightarrow \ell^{+} \ell^{-} \ell^{\prime +} \ell^{\prime -} + {\rm jet} + X$ at the $14~ {\rm TeV}$ LHC. }
\label{LO-pTl}
\end{figure}
\begin{figure}[htbp]
\begin{center}
\includegraphics[width=10cm]{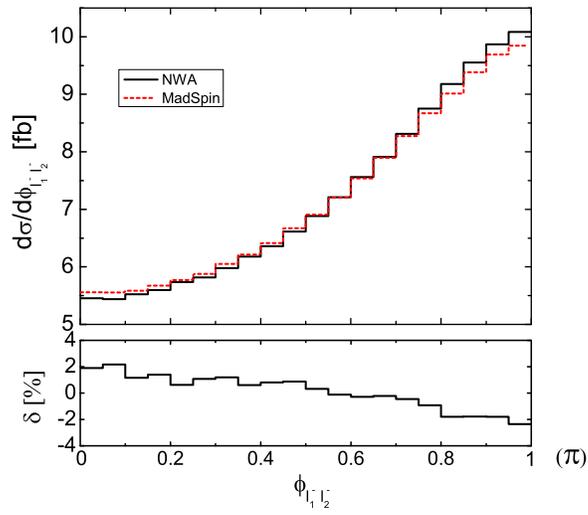}
\caption{ The LO distributions of the azimuthal angle between the two negatively charged leptons for $pp \rightarrow ZZ + {\rm jet} \rightarrow \ell^{+} \ell^{-} \ell^{\prime +} \ell^{\prime -} + {\rm jet} + X$ at the $14~ {\rm TeV}$ LHC.}
\label{LO-phi}
\end{center}
\end{figure}

First, we present the LO distributions by applying both the naive narrow width approximation (NWA) and MadSpin methods in Figs.\ref{LO-pTl}(a,b) and Fig.\ref{LO-phi}, to demonstrate the spin correlation and finite width effects from the $Z$-boson leptonic decays. The relative deviation is defined as
\begin{eqnarray}
\delta(x)
=
\Big( \frac{d\sigma_{{\rm MadSpin}}}{dx} - \frac{d\sigma_{{\rm NWA}}}{dx} \Big)\Big/\frac{d\sigma_{{\rm NWA}}}{dx},
\end{eqnarray}
where $x = p_{T, \ell_1^-}$, $x = p_{T, \ell_1^-}$ and $\phi_{\ell_1^- \ell_2^-}$. As shown in Fig.\ref{LO-pTl}(a), the transverse momentum distribution of the leading lepton is enhanced by the spin correlation and finite width effects when $p_{T, \ell_1^-} < 50~ {\rm GeV}$, while is suppressed in the region of $p_{T, \ell_1^-} \in [50, 190]~ {\rm GeV}$, compared to the one obtained by using the naive NWA method. Correspondingly, the relative deviation can reach $6.21\%$ at $p_{T, \ell_1^-} \sim 30~ {\rm GeV}$ and $-2.84\%$ at $p_{T, \ell_1^-} \sim 90~ {\rm GeV}$ in the plotted $p_{T, \ell_1^-}$ region. From Fig.\ref{LO-pTl}(b) we see that the spin correlation and finite width effects in the transverse momentum distribution of the second lepton are more apparent, and the relative deviation varies between $4.75\%$ and $-14.45\%$ for $p_{T,\ell_2^-}$ in the range of $[0, 140]~ {\rm GeV}$. In analogy to the transverse momentum distributions of the final leptons, the distributions of the azimuthal-angle separation of the two negatively charged leptons depicted in Fig.\ref{LO-phi} also demonstrate sizable spin correlation and finite width effects. The corresponding relative deviation varies from $2.17\%$ to $-2.36\%$ in the plotted $\phi_{\ell_1^- \ell_2^-}$ region. We can conclude from Figs.\ref{LO-pTl}(a,b) and Fig.\ref{LO-phi} that the off-shell contribution and spin correlation from the $Z$-boson leptonic decays are nonnegligible, and therefore should be considered in NLO QCD+EW precision calculation.

\par
\begin{figure}[htbp]
\includegraphics[width=10cm]{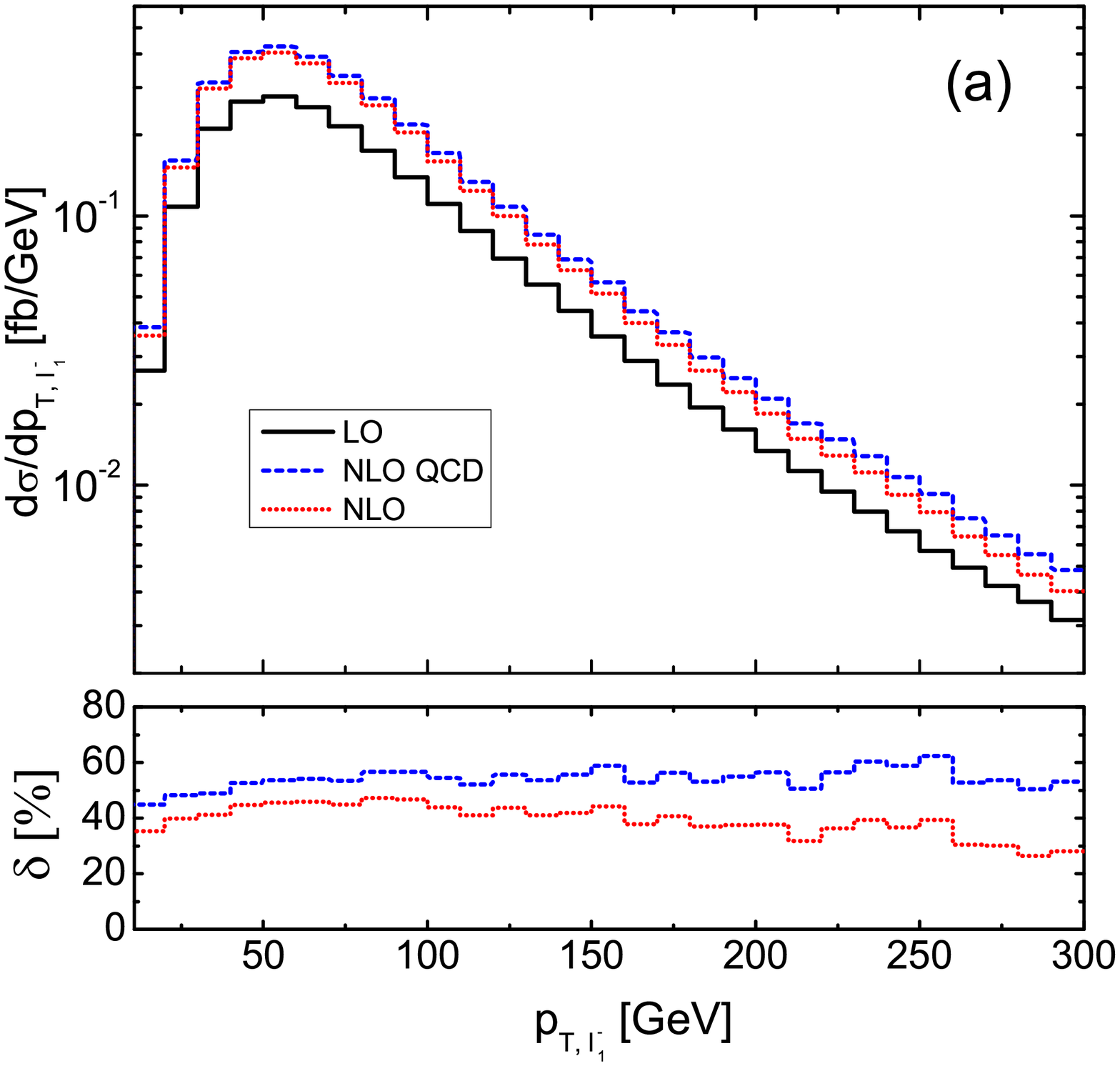}
\hspace{-2cm}
\includegraphics[width=10cm]{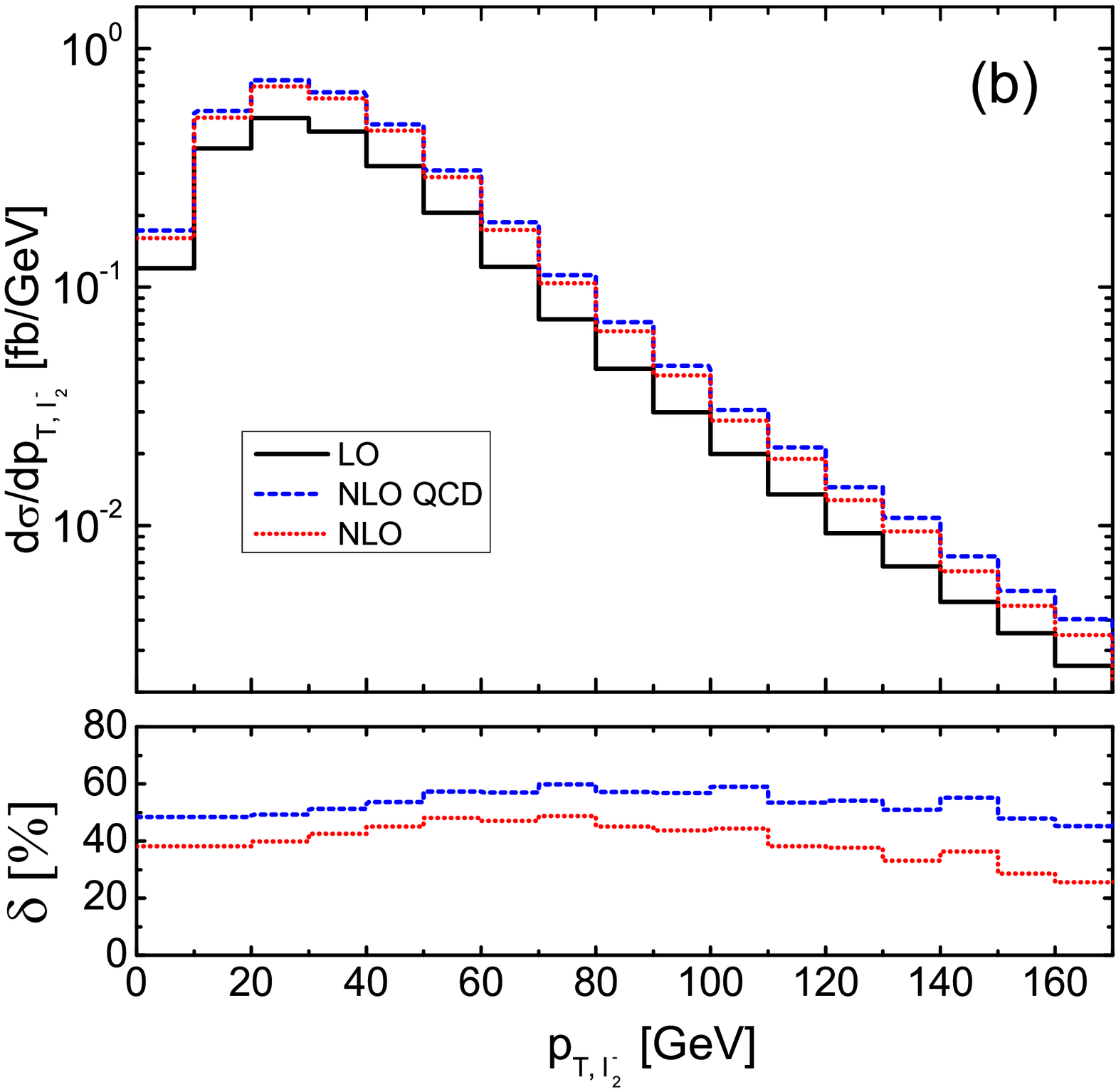}
\caption{ The LO, NLO QCD and NLO QCD+EW corrected transverse momentum distributions of (a) leading and (b) second leptons for $pp \rightarrow ZZ + {\rm jet} \rightarrow \ell^{+} \ell^{-} \ell^{\prime +} \ell^{\prime -} + {\rm jet} + X$ at the $14~ {\rm TeV}$ LHC.}
\label{NLO-pTl}
\end{figure}

In Figs.\ref{NLO-pTl}(a) and (b) we depict the LO, NLO QCD and NLO QCD+EW corrected transverse momentum distributions of the leading and second negatively charged leptons for $pp \rightarrow ZZ + {\rm jet} \rightarrow \ell^{+} \ell^{-} \ell^{\prime +} \ell^{\prime -} + {\rm jet} + X$ by adopting the MadSpin method. The corresponding relative corrections are shown in the nether plots. We see from the figures that both the LO and NLO corrected $p_{T,\ell_1^-}$ distributions reach their maxima at $p_{T,\ell_1^-} \sim 60~ {\rm GeV}$, while the $p_{T,\ell_2^-}$ distributions peak at $p_{T,\ell_2^-} \sim 30~ {\rm GeV}$. The relative corrections for the transverse momentum distribution of $\ell_1^-$ exhibit similar behavior with $\ell_2^-$. The relative QCD correction is fairly stable in the whole plotted $p_T$ range. In contrast, the relative EW correction becomes significant in high $p_T$ region, of about $-16.3\%$ at $p_{T,\ell_1^-} \sim 300~ {\rm GeV}$ for the leading lepton and $-14.1\%$ at $p_{T,\ell_2^-} \sim 170~ {\rm GeV}$ for the second lepton.

\par
\begin{figure}[htbp]
\begin{center}
\includegraphics[width=10cm]{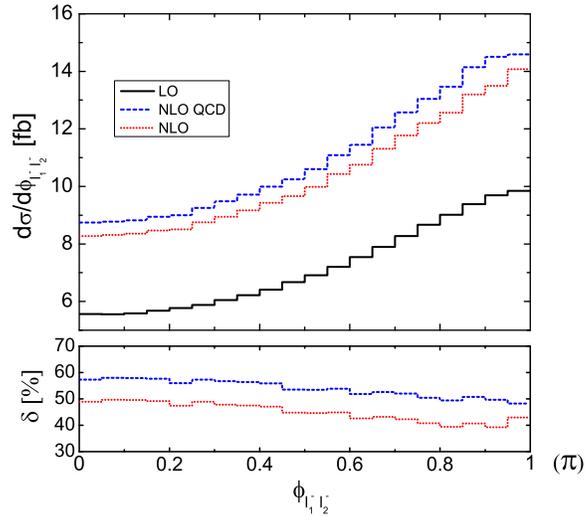}
\caption{ The LO, NLO QCD and NLO QCD+EW corrected distributions of the azimuthal angle between the two negatively charged leptons for $pp \rightarrow ZZ + {\rm jet} \rightarrow \ell^{+} \ell^{-} \ell^{\prime +} \ell^{\prime -} + {\rm jet} + X$ at the $14~ {\rm TeV}$ LHC. }
\label{NLO-phi}
\end{center}
\end{figure}

In  Fig.\ref{NLO-phi} we present the LO, NLO QCD, NLO QCD+EW corrected distributions of the azimuthal angle between the two negatively charged leptons and the corresponding relative corrections for $pp \rightarrow ZZ + {\rm jet} \rightarrow \ell^{+} \ell^{-} \ell^{\prime +} \ell^{\prime -} + {\rm jet} + X$ by employing the MadSpin method. From the figure we see clearly that the two negatively (as well as positively) charged leptons in final state prefer to be back-to-back. The NLO QCD correction enhances the LO $\phi_{\ell_1^- \ell_2^-}$ distribution remarkably, and the relative QCD correction decreases from $57.3\%$ to $48.2\%$ with the increment of $\phi_{\ell_1^- \ell_2^-}$ from $0$ to $\pi$. While the NLO EW correction suppresses the LO $\phi_{\ell_1^- \ell_2^-}$ distribution slightly, and the relative EW correction is much stable, varying in the range of $[-6.96\%, -5.24\%]$. Consequently, the full NLO relative correction varies from $49.7\%$ to $39.3\%$ in the plotted $\phi_{\ell_1^- \ell_2^-}$ region.

\vskip 5mm
\section{Summary}  \label{summary}
\par
In this paper, we calculate the NLO QCD + NLO EW corrections to the $ZZ+{\rm jet}$ production including subsequent $Z$-boson leptonic decays at the $14~{\rm TeV}$ LHC. In dealing the $Z$-boson leptonic decays, we employ the MadSpin method to take into account the spin correlation and finite width effects. Our numerical results show that the off-shell contribution and spin correlation from the $Z$-boson leptonic decays should be included in precision calculation. The NLO EW correction is relatively small compared to the NLO QCD correction, but is nonnegligible for precision theoretical predictions, particularly in high transverse momentum and invariant mass regions due to the Sudakov effect. Our analysis of the factorization/renormalization scale dependence of the integrated cross section affirm that the NLO QCD+EW correction can significantly reduce the scale uncertainty.

\vskip 5mm
\section{Acknowledgments}
This work was supported in part by the National Natural Science Foundation of China (Grant No. 11275190, No. 11375171, No. 11405173, No. 11535002, No. 11375008).

\end{document}